\documentclass[letterpaper,12pt]{article}
\usepackage[margin=0.75in]{geometry}
\usepackage[T1]{fontenc}
\usepackage{graphicx, lipsum, textcomp, float} 
\usepackage{hyperref} 
\usepackage{advdate, datenumber} 
\usepackage{amsmath} 
\usepackage{listings} 
\usepackage{multirow}
\usepackage{comment}
\usepackage{xcolor}

\usepackage{fontawesome5}
\newcommand{\orcid}[1]{\href{https://orcid.org/#1}{\textcolor[HTML]{A6CE39}{\faOrcid}}}
\newcommand{\github}[1]{\href{https://github.com/#1}{\textcolor[HTML]{70726E}{\faGithub}}}

\usepackage[]{lineno}

\usepackage[scaled]{beramono}
 
\definecolor{darkgreen}{rgb}{0.05, 0.3, 0.1}
\let\oldtexttt\texttt
\renewcommand{\texttt}[1]{\oldtexttt{\textcolor{darkgreen}{#1}}}

\usepackage{setspace}
\usepackage[
    style=ieee, 
    isbn=false, 
    url=true, 
    natbib=true, 
    backend=bibtex,
    maxcitenames=1,
    mincitenames=1
    ]{biblatex}

\usepackage{caption}
\usepackage{tocbasic}
\DeclareTOCStyleEntry[
  beforeskip=.2em plus 1pt,
  pagenumberformat=\textbf
]{tocline}{section}

\usepackage{abstract}
\usepackage{titling}

\definecolor{mypurple}{RGB}{140,54,140}
\definecolor{homered}{RGB}{127, 0, 10}
\definecolor{officeorange}{RGB}{204, 75, 0}
\definecolor{mauroblue}{RGB}{53, 48, 217}
\definecolor{citegreen}{RGB}{15, 133, 13}
\definecolor{hyperlinkpurple}{RGB}{42, 0, 163}
\hypersetup{
    colorlinks=true,
    linkcolor=hyperlinkpurple,
    filecolor=mypurple,      
    urlcolor=teal,
    citecolor=citegreen
}

\title{
\textbf{
Efficient Structure-Informed Featurization and Property Prediction of Ordered, Dilute, and Random Atomic Structures}
}

\author{
\textbf{Adam M. Krajewski\textsuperscript{a}}
\orcid{0000-0002-2266-0099} \github{amkrajewski}, 
Jonathan W. Siegel \textsuperscript{b}
\orcid{0000-0002-1493-4889} \github{jwsiegel2510}, 
Zi-Kui Liu\textsuperscript{a}
\orcid{0000-0003-3346-3696}
\\
\footnotesize{
a. Department of Materials Science and Engineering, The Pennsylvania State University, USA}\\
\footnotesize{b. Department of Mathematics, Texas A\&M University, USA}\\
\footnotesize{Corresponding Author: A. M. Krajewski, adam@phaseslab.org | ak@psu.edu}
}

\renewcommand{\maketitlehookd}{%
\begin{abstract}

Structure-informed materials informatics is a rapidly evolving discipline of materials science relying on the featurization of atomic structures or configurations to construct vector, voxel, graph, graphlet, and other representations useful for machine learning prediction of properties, fingerprinting, and generative design. This work discusses how current featurizers typically perform redundant calculations and how their efficiency could be improved by considering (1) fundamentals of crystallographic (orbits) equivalency to optimize ordered cases and (2) representation-dependent equivalency to optimize cases of dilute, doped, and defect structures with broken symmetry. It also discusses and contrasts ways of (3) approximating random solid solutions occupying arbitrary lattices under such representations.

Efficiency improvements discussed in this work were implemented within \texttt{pySIPFENN} or \textit{python toolset for Structure-Informed Property and Feature Engineering with Neural Networks} developed by authors since 2019 and shown to increase performance from 2 to 10 times for typical inputs. Throughout this work, the authors explicitly discuss how these advances can be applied to different kinds of similar tools in the community.

\rule{0.88\textwidth}{0.5pt}
\vspace{48pt}
\end{abstract}
}

\begin{document}

\maketitle

\setstretch{1}

\section*{Highlights}
\begin{itemize}
    \item \texttt{pySIPFENN} is an open toolset for Structure-Informed Property and Feature Engineering
    \item Modular build enables both easy extensions and integration into external libraries
    \item \texttt{KS2022} featurizer multiplies machine learning throughput using symmetry-equivalency
    \item \texttt{KS2022\_dilute} enables further optimizations by considering representation-equivalency
    \item \texttt{KS2022\_randomSolutions} considers chemistry and geometry to featurize solid solutions
\end{itemize}

\tableofcontents

\setstretch{1.5}



\section{Introduction} \label{sec:Introduction}

\texttt{SIPFENN} or \textit{Structure-Informed Prediction of Formation Energy using Neural Networks} software, first introduced by the authors in 2020 \cite{Krajewski2022ExtensibleNetworks, Krajewski2020SIPFENNModels}, is one of several open-source tools available in the literature \cite{Ward2017, Jha2019IRNet, Chen2019GraphCrystals, Choudhary2021AtomisticPredictions, Deng2023CHGNetModelling, Davariashtiyani2023FormationRepresentation, Davariashtiyani2023FormationRepresentation, Schmidt2023Machine-Learning-AssistedMaterials} which train machine learning (ML) models on the data from large Density Functional Theory (DFT) based datasets like \texttt{OQMD} \cite{Saal2013MaterialsOQMD, Kirklin2015TheEnergies, Shen2022ReflectionsOQMD}, \texttt{AFLOW} \cite{Curtarolo2013AFLOW:Discovery, Toher2018TheDiscovery}, Materials Project \cite{Jain2013Commentary:Innovation}, NIST-\texttt{JARVIS}\cite{Choudhary2020TheDesign}, \texttt{Alexandria} \cite{Schmidt2022AFunctionals}, or \texttt{GNoME} \cite{Merchant2023ScalingDiscovery} to predict formation energies of arbitrary atomic structures, with accuracy high enough to act as a low-cost surrogate in the prediction of thermodynamic stability of ground and non-ground state configurations at 0K temperature. The low runtime cost allows such models to efficiently screen through millions of different atomic structures of interest on a personal machine in a reasonable time. 

In addition to high-accuracy neural network models trained on \texttt{OQMD} \cite{Saal2013MaterialsOQMD, Kirklin2015TheEnergies, Shen2022ReflectionsOQMD}, \texttt{SIPFENN} included a number of features not found in competing tools available at the time, such as the ability to quickly readjust models to a new chemical system based on just a few DFT data points through transfer learning and a selection of models optimized for different objectives like extrapolation to new materials instead of overfitting to high-data-density regions or low memory footprint \cite{Krajewski2022ExtensibleNetworks}.

\texttt{SIPFENN}'s usefulness has been demonstrated, for instance, in the cases where the structure of an experimentally observed compound could not be identified in industry-relevant Nd-Bi \cite{Im2022ThermodynamicModeling} and Al-Fe \cite{Shang2021FormingJoints} systems and had to be predicted. This was accomplished by (1) high-throughput generation of hundreds of thousands of possible candidates with the exact stoichiometry based on elemental substitutions into structures from both open DFT-based databases \cite{Saal2013MaterialsOQMD, Kirklin2015TheEnergies, Shen2022ReflectionsOQMD, Curtarolo2013AFLOW:Discovery, Toher2018TheDiscovery, Jain2013Commentary:Innovation, Choudhary2020TheDesign, Schmidt2022AFunctionals, Merchant2023ScalingDiscovery} and experimentally observed ones from Crystallography Open Database (COD) \cite{Grazulis2009CrystallographyStructures, Grazulis2012CrystallographyCollaboration, Grazulis2019CrystallographyPerspectives}, followed by (2) selection of thousands of low-energy candidates, (3) down-selection of tens of unique candidates based on clustering in the \texttt{SIPFENN}'s feature space, and (4) final validation with DFT and experiments. It has also been deployed in several thermodynamic modeling studies, e.g. of Nb-Ni system \cite{Sun2023ThermodynamicESPEI}, in conjunction with DFT and experimental data processed through \texttt{ESPEI} \cite{Bocklund2019ESPEICuMg} to automatically fit parameters of CALPHAD \cite{Olson2023GenomicDynamics} models deployed in \texttt{pycalphad} \cite{Otis2017Pycalphad:Python}.

\section{General Structure Featurization Improvements} \label{sec:featurization}

\subsection{pySIPFENN Overview and Core Advantages} \label{ssec:coreimprovements}

Being able to predict the thermodynamic stability of arbitrary atomic structures and their modifications is one of the most critical steps in establishing whether hypothetical candidates can be made in real life \cite{Zunger2019BewareMaterials}; however, it is certainly not the only task of interest to the community \cite{Jha2023MachineChallenges, Tao2021MachineDiscovery}. These diverse needs, combined with increasing interest in multi-property modeling, have shifted the focus of \texttt{SIPFENN} tool from model training \cite{Krajewski2022ExtensibleNetworks} toward the development of reliable, easy-to-use, and efficient general-purpose featurizers existing in a framework, which can be used by researchers and companies to quickly develop and deploy property-specific models, or use features directly in exploring similarity and trends in materials.

Thus, while the acronym has been retained, the name of the software has been changed to \textit{python toolset for Structure-Informed Property and Feature Engineering with Neural Networks} or \texttt{pySIPFENN}, and the software component has been carefully re-implemented in its entirety to make it as general as possible and enable the following core advantages:

\begin{enumerate}
    
    \item Reliable featurization, which can be immediately transferred to other tools thanks to standalone submodule implementations based only on two common libraries (\texttt{NumPy} \cite{Harris2020ArrayNumPy} and \texttt{pymatgen} \cite{Ong2013PythonAnalysis}). These include completely re-written \texttt{Ward2017} Java-based featurizer \cite{Ward2017} (see Section \ref{ssec:Ward2017Translation}) and 3 new ones, described in Sections \ref{sec:ordered}, \ref{sec:dilute}, and \ref{sec:randomsolutions}.

    \item Effortless plug-and-play deployment of neural network (and other) ML models (for any property) utilizing any of the defined feature vectors, enabled by the use of Open Neural Network Exchange (\texttt{ONNX}) open-source format \cite{Bai2019ONNX:Exchange} which can be exported from nearly every modern ML framework and is then loaded into \texttt{pySIPFENN}'s \texttt{PyTorch} backend \cite{Paszke2019PyTorch:Library} through \texttt{onnx2torch} \cite{Kalgin2021Onnx2torch:PyTorch}. Furthermore, implementing custom predictors, beyond those supported by \texttt{PyTorch}, is made easy by design.

    \item Dedicated \texttt{ModelExporters} submodule makes it easy to export trained models for publication or deployment on a different device while also enabling weight quantization and model graph optimizations to reduce memory requirements.

    \item The ability to acquire data and adjust or completely retrain model weights through automated \texttt{ModelAdjusters} submodule. Its applications include:
    \begin{enumerate}
        \item Fine-tuning models based on additional local data to facilitate transfer learning ML schemes of the domain adaptation kind \cite{Ben-David2010ADomains}, where a model can be adjusted to new chemistry and specific calculation settings, introduced by \texttt{SIPFENN} back in 2020 \cite{Krajewski2022ExtensibleNetworks}, which is also being adopted by other models like \texttt{ALIGNN} \cite{Gupta2024Structure-awareDatasets}. Such an approach can also be used iteratively in active learning schemes where new data is obtained and added.
        
        \item Tuning or retraining of the models based on community atomistic databases, or their subsets, accessed through \texttt{OPTIMADE API} queried by \texttt{optimade-python-tools} library \cite{Andersen2021OPTIMADEData, Evans2024DevelopmentsExchange, Evans2021optimade-python-tools:APIs}, to adjust the model to a different domain, which in the context of DFT datasets could mean adjusting the model to predict properties with DFT settings used by that database or focusing its attention to specific chemistry like, for instance, all compounds of Sn and all perovskites. Critically, this functionality requires user to only provide a standard human-readable \texttt{OPTIMADE} query like the one below.\\
        \begin{center}
        \texttt{'elements HAS "Hf" AND elements HAS "Mo" }\\
        \texttt{AND NOT elements HAS ANY "O","C","F","Cl","S"'}
        \end{center}
        
        \item Knowledge transfer learning \cite{Torrey2010HandbookLearning} to adjust models to entirely new, often less available properties while harvesting the existing pattern recognition.
    \end{enumerate}    
\end{enumerate}

\begin{figure}[h!]
    \centering
    \includegraphics[width=0.85\textwidth]{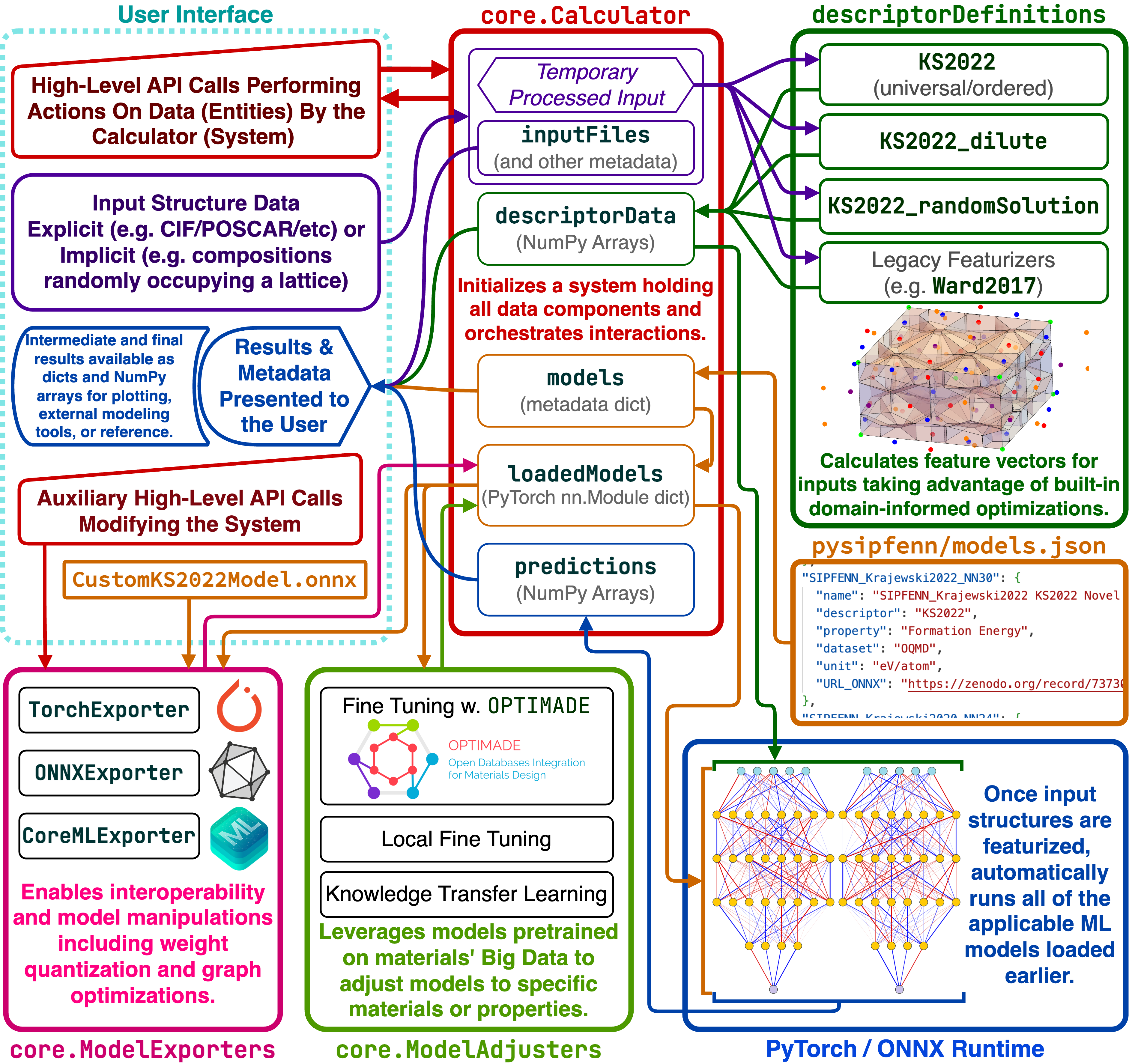}
    \caption{Main schematic of \texttt{pySIPFENN} framework detailing the interplay of internal components described in Section \ref{ssec:coreimprovements}. The user interface provides a high-level API to process structural data within \texttt{core.Calculator}, pass it to featurization submodules in \texttt{descriptorDefinitions} to obtain vector representation, then passed to models defined in \texttt{models.json} and (typically) run automatically through all available models. All internal data of \texttt{core.Calculator} is accessible directly, enabling rapid customization. An auxiliary high-level API enables advanced users to operate and retrain the models.}
    \label{fig:pySIPFENNMainSchematic}
\end{figure}

The resulting \texttt{pySIPFENN} computational framework is composed of several components, as depicted in Figure~\ref{fig:pySIPFENNMainSchematic}, and is available through several means described in Section \ref{sec:softwareavaialbility}, alongside high-quality documentation and examples.

\subsection{Ward2017 Reimplementation} \label{ssec:Ward2017Translation}

In their 2017 work \citet{Ward2017} introduced a novel atomic structure featurization concept based on establishing and weighting neighbor interactions by faces from 3D Voronoi tesselation to describe local chemical environments (LCEs) of atomic sites and then performing statistics over them to obtain a global feature vector. The original \texttt{SIPFENN} models \cite{Krajewski2020SIPFENNModels} built on top of this while utilizing an improved, carefully designed deep neural network models to obtain up to 2.5 times lower prediction error on the same dataset \cite{Krajewski2022ExtensibleNetworks}. A detailed description of the descriptor can be found in Section 2.1 of \citet{Krajewski2022ExtensibleNetworks}. In general, the calculation of the \texttt{Ward2017} descriptor consists of three parts:

\begin{itemize}
    \item Calculation of attributes based upon global averages over the components of the structure.
    \item Calculation of attributes based upon local neighborhood averages for each site in the structure.
    \item Calculation of more complex attributes based upon averages over paths in the structure.
\end{itemize}

\citet{Ward2017} implemented the above calculations in Java, which was popular at the time; while most of the current machine-learning packages use almost exclusively Python (e.g., \texttt{scikit-learn} \cite{PedregosaFABIANPEDREGOSA2011Scikit-learn:Python} and \texttt{PyTorch} \cite{Paszke2019PyTorch:Library}), making it cumbersome to use Java. Even more critically, the original Java implementation was not computationally efficient (as explored in Sections \ref{sec:ordered}, \ref{sec:dilute}, and \ref{sec:randomsolutions}), and enabling tools were not supported in Java.

In the present work, authors have reimplemented \citet{Ward2017} from scratch in Python as a standalone submodule for \texttt{pySIPFENN}, which calculates all 271 features within numerical precision, except for three performing a random walk on the structure, which is stochastic in nature and results in slightly different final values due to a different seed. The Voronoi tessellation has been implemented with \texttt{Voro++} \cite{Rycroft2007MultiscaleFlow, Rycroft2009Voro++:C++, Lu2023AnCells} and all numerical operations were written using \texttt{NumPy} \cite{Harris2020ArrayNumPy} arrays to greatly speed up the calculations and make the efficient utilization of different computing resources, such as GPUs, easy to implement.

\subsection{KS2022 Feature Optimization} \label{ssec:ks2022features}

Typically, during feature engineering, researchers first attempt to collect all features expected to enable good inference and then remove some based on the interplay of several factors:
\begin{enumerate}
    \item \textbf{Low impact} on the performance, which increases the representation memory requirements and possibly increases the risk of overfitting to both systematic and random trends. 
    \label{item:featureoptimize1}
    \item \textbf{High computational cost}, which limits the throughput of the method deployment.
    \label{item:featureoptimize2}
    \item \textbf{Unphysical features or feature representations} which can improve model performance against well-behaving benchmarks covering a small subset of the problem domain but compromise model interpretability and extrapolation ability in unpredictable ways.
    \label{item:featureoptimize3}
\end{enumerate}

The \texttt{KS2022} feature set, added in \texttt{pySIPFENN v0.10} in November 2022, is a significant modification of the \texttt{Ward2017} \cite{Ward2017}, which focuses on points \ref{item:featureoptimize2} and \ref{item:featureoptimize3} above while enabling optimizations described in Sections \ref{sec:ordered} through \ref{sec:randomsolutions} and delegating the removal of low-impact features to modeling efforts and keeping featurization as problem-independent as possible.

First, all 11 features relying on representation of crystal symmetry space groups with space group number \texttt{float}s rather than classes (e.g. using one-hot vectors) have been removed due to the unphysical nature of such representation leading to, for instance, BCC ($229$) being much closer to FCC ($225$) than to just slightly uniaxially distorted BCC ($139$), which itself would be very close to trigonal structures. 

Next, featurization code has been thoroughly profiled in regard to time spent on the execution of feature-specific subroutines and analyzed in the context of feature importance identified in the past work \cite{Krajewski2022ExtensibleNetworks}. This led to the removal of the 1 \textit{CanFormIonic} feature, which relied on combinatorically expensive guessing of oxidation states, and 3 features based on Warren-Cowley (WC) parameters \cite{Cowley1950AnAlloys}, which were relatively very expensive without significantly contributing to the performance due to scarcity of disordered structures in most atomistic datasets. However, the authors intend to add them back in future problem-specific feature sets using a recently released high-performance library by \citet{Gehringer2023ModelsSimple}. 

Together, 15 features were removed, bringing the total number of the \texttt{KS2022} features to $256$ while disproportionately improving the featurization speed. For instance, in the case of featurization of 30 sites in a disordered (no symmetry) structure, \texttt{KS2022} is $2.3$ times faster than \texttt{Ward2017} ($430$ms vs $990$ms single-threaded on Apple M2 Max).

\section{Optimizations for Ordered Structures} \label{sec:ordered}

Modeling of disordered materials is a critical area of research \cite{Zaki2023Glassomics:Intelligence}; however, the vast majority of atomistic ab initio datasets used for ML studies focus on highly symmetric ordered structures because of their high availability and ability to model complex phenomena in a holistic fashion if local ergodicity can be established \cite{Liu2022TheoryTheorem, Liu2023ThermodynamicsPerspectives}. One evidence of the focus on such data is the fact that out of $4.4$ million atomic structures in \texttt{MPDD} \cite{Krajewski2021MPDD:Database}, which includes both DFT-based \cite{Saal2013MaterialsOQMD, Kirklin2015TheEnergies, Shen2022ReflectionsOQMD, Curtarolo2013AFLOW:Discovery, Toher2018TheDiscovery, Jain2013Commentary:Innovation, Choudhary2020TheDesign, Merchant2023ScalingDiscovery} and experimental \cite{Grazulis2009CrystallographyStructures, Grazulis2012CrystallographyCollaboration, Grazulis2019CrystallographyPerspectives} data, only $54$ thousand or $1.25\%$ lack any symmetry. It is also worth noting that this number used to be much lower before the recent publication of the \texttt{GNoME} dataset by Google DeepMind \cite{Merchant2023ScalingDiscovery}, which accounts for around $\frac{3}{4}$ of them. 

In the case of remaining $98.75\%$ structures, a 3-dimensional crystallographic spacegroup is defined for each of them along with corresponding \emph{Wyckoff positions} (designated by letters) which are populated with either zero (empty), one (when symmetry-fixed), or up to infinitely many (typically up to a few) atoms forming a set of symmetry-equivalent sites called \emph{crystallographic orbits} \cite{Muller2006RemarksPositions}. When these crystallographic orbits are collapsed into atoms occupying a unit cell, each is repeated 
based on the \emph{multiplicity} associated with the Wyckoff position it occupies, which can range from 1 up to 192 (e.g., position l in Fm-3m/225), with values 1, 2, 3, 4, 6, 8, 16, 24, 32, 48, and 96 being typical \cite{Mehl2016ThePrototypes} even in compositionally simple materials like one of the experimentally observed allotropes of pure silicon with atoms at the 8a, 32e, and 96g positions \cite{Gryko2000Low-densityGap}. For certain crystal lattice types, the multiplicity can be somewhat reduced by redefining their spatial periodicity with so-called \textit{primitive} unit cells, like in the case of the aforementioned Si allotrope, in which primitive unit cell has 4 times fewer (34) sites but still over 10 times more than the 3 unique crystallographic orbits.

\begin{figure}[h]
    \centering
    \includegraphics[width=0.98\textwidth]{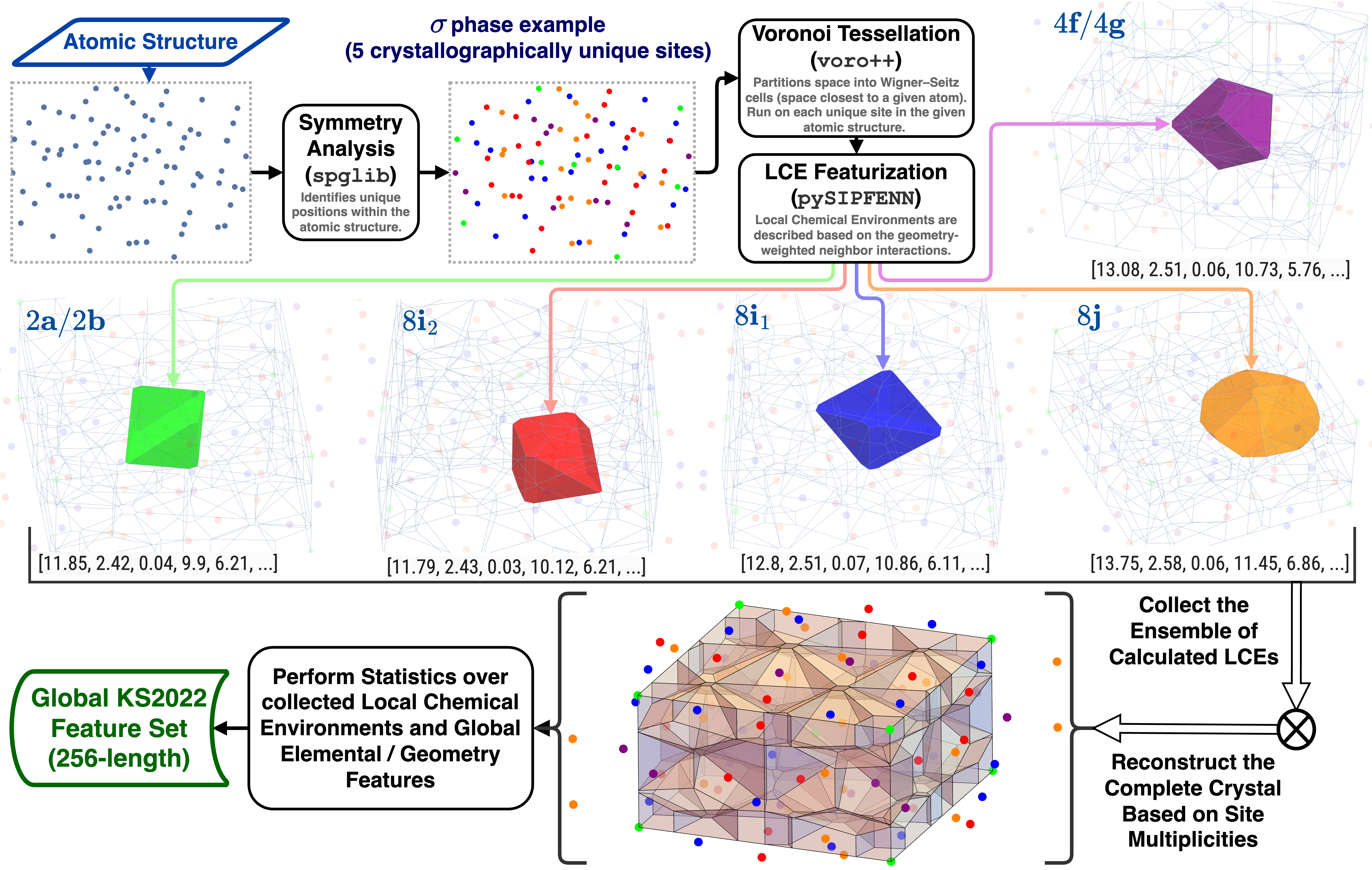}
    \caption{
    Schematic of the general-purpose \texttt{KS2022} featurization routine with built-in optimization for ordered structures. First, the atomic structure (in \texttt{pymatgen Structure} object format \cite{Ong2013PythonAnalysis}) is loaded, and sites in it are annotated with their crystallographic orbits using \texttt{spglib} \cite{Togo2018Spglib:Search}. Then, one site is selected from each orbit to form a set of unique sites, for which Wigner-Seitz cells (depicted as colored polyhedra) are calculated with \texttt{Voro++} \cite{Rycroft2007MultiscaleFlow, Rycroft2009Voro++:C++, Lu2023AnCells} and featurized to get site-specific local chemical environment (LCE) descriptors. The complete site ensemble is then reconstructed based on multiplicities of Wyckoff positions corresponding to the sites. A non-trivial example of $\sigma$-phase with 30 atoms belonging to 5 crystallographic orbits with interesting Wigner-Seitz cells (relative to usually shown FCC/BCC ones \cite{Bohm1996VoronoiLattices}) has been depicted.
    }
    \label{fig:ks2022}
\end{figure}

This presents an immediate and previously untapped opportunity for multiplying the computational performance of most atomistic featurizers (e.g., \texttt{Matminer} \cite{Ward2018Matminer:Mining}) and ML models \cite{Ward2017, Chen2019GraphCrystals, Jha2019IRNet, Krajewski2022ExtensibleNetworks, Choudhary2021AtomisticPredictions, Deng2023CHGNetModelling, Davariashtiyani2023FormationRepresentation, Schmidt2023Machine-Learning-AssistedMaterials, Banik2024EvaluatingMaterials, Hu2021Atomtransmachine:Learning}, which nearly always process all atoms given in the input structure occasionally converting to primitive unit cell in certain routines (\texttt{CHGNet} \cite{Deng2023CHGNetModelling}), unless they operate on different occupancies of the same structure \cite{Crivello2022SupervisedExample}. This allows for a proportional decrease in both CPU/GPU time and memory footprint. The general-purpose \texttt{KS2022} in \texttt{pySIPFENN} uses high-performance symmetry analysis library \texttt{spglib} \cite{Togo2018Spglib:Search} to automatically take advantage of this whenever possible, as depicted in the schematic in Figure~\ref{fig:ks2022}. It shows an interesting example of a topologically close-packed $\sigma$ phase, which is critical to model in a wide range of metallic alloys \cite{Joubert2008CrystalPhase} but challenging in terms of combinatorics because of 5 unique sites that can be occupied by many elements \cite{Choi2019ADesign, Ostrowska2020ThermodynamicW} making it a very active area of ML modeling efforts \cite{Crivello2022SupervisedExample, Zha2024ApplyingEnergy} in the thermodynamics community.

In the case of \texttt{KS2022} featurizer, running the same 30-atom test as in Section \ref{ssec:ks2022features} but on $\sigma$ phase takes on average $84$ms or is 5.1 times faster thanks to processing 6 times less sites. Similar results should be (a) quickly achievable with any other featurizer processing individual sites, including most graph representations embedding local environments (e.g., \texttt{MEGNet} \cite{Chen2019GraphCrystals}) or deconstructing graphs into graphlets (e.g., \texttt{minervachem} molecule featurizer \cite{Tynes2024LinearCheminformatics}), and (b) possible with convolution-based models operating on graphs (e.g., \texttt{ALIGNN} \cite{Choudhary2021AtomisticPredictions}) or voxels \cite{Davariashtiyani2023FormationRepresentation} through custom adjustments to the specific convolution implementation. In the case of voxel representations and any other memory-intense ones, it may also be beneficial to utilize this approach to compress them when transferring between devices like CPU and GPU or across an HPC network.

\section{Optimizations for Dilute, Defect, and Doped Structures} \label{sec:dilute}

The optimization strategy in Section \ref{sec:ordered} ensures that only the sites that are \emph{guaranteed} to be \emph{crystallographically unique} are processed through featurization or graph convolution and is directly applicable to the vast majority of both data points and literature methods. However, in the case of methods relying on describing the immediate neighbors, whether through Wigner-Seitz cell (see Fig. \ref{fig:ks2022}) or subgraph (see, e.g., \cite{Chen2019GraphCrystals}), one can achieve further efficiency improvements by considering which sites are \emph{guaranteed} to be \emph{unique under the representation}.

There are several classes of atomic structures where the distinction above makes a difference, but the room to improve is exceptionally high when one site (or otherwise small subset) in an otherwise highly symmetric structure is modified, leading to a structure that, depending on the context, will be typically called \emph{dilute} when discussing alloys \cite{Chong2021CorrelationAlloys}, \emph{doped} when discussing electronic materials \cite{Chen2022InteractionStudy}, or said to have \emph{defect} in a more general sense \cite{Castleton2009DensitySupercells}. Throughout \texttt{pySIPFENN}'s codebase and the rest of this work, the single term \emph{dilute} is used to refer to all of such structures due to authors' research focus on Ni-based superalloys at the time when optimizations below were open-sourced in February 2023.

To visualize the concept, one can consider, for instance, a 3x3x3 body-centered cubic (BCC) conventional supercell (54 sites) and call it \textit{base structure}. If it only contains a single specie, then \texttt{KS2022} from Section \ref{sec:ordered} will recognize that there is only one crystallographic orbit and only process that one. However, if a substitution is made at any of the 54 equivalent sites, the space group will change from Im-3m (229) to Pm-3m (221), with 8 crystallographic orbits on 7 Wyckoff positions; thus, the default \texttt{KS2022} featurizer will process 8 sites. 

\begin{figure}[h]
    \centering
    \includegraphics[width=0.85\textwidth]{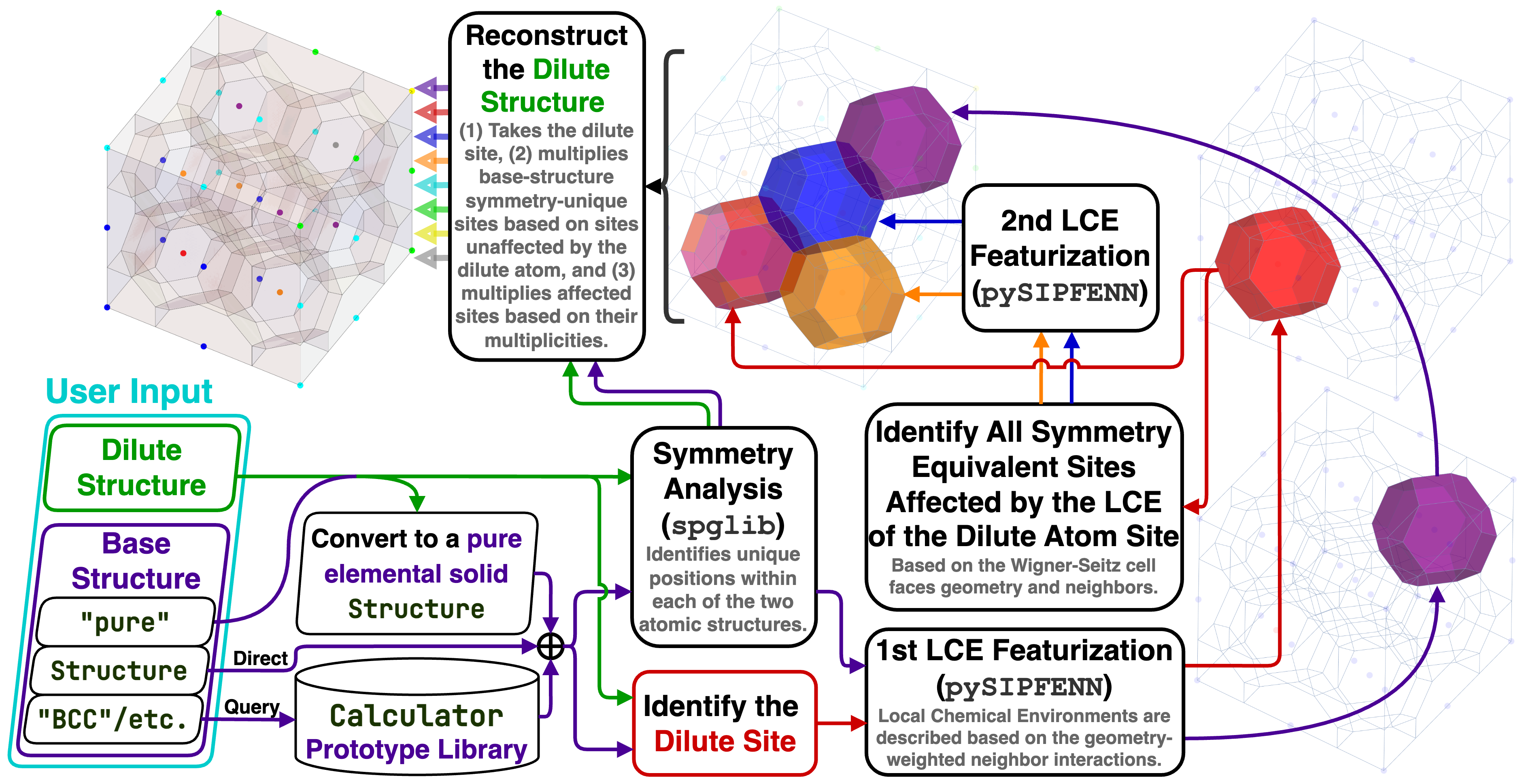}
    \caption{Core schematic of the \texttt{KS2022\_dilute} featurizer. The dilute structure is compared to either the explicit or implicit base structure to identify the dilute site, which is then featurized alongside all crystallographically unique sites in the base structure. Information extracted from dilute structure featurization is then used to identify previously-equivalent sites affected by it, which go through the second round of featurization. Lastly, the complete ensemble is reconstructed, and \texttt{KS2022} are obtained. BCC supercell is used as an example.}
    \label{fig:KS2022dilute}
\end{figure}

At the same time, several of these crystallographic orbits will be differentiated \emph{only} by the orientation and distance to the dilute (substitution) site, which \emph{does} affect ab initio calculation results (e.g., vacancy formation energy vs supercell size \cite{Hargather2022ANi}), but is \emph{guaranteed} to \emph{have no effect on the model's representation} because of the exact same neighborhood configuration (including angles and bond lengths) if conditions given earlier are met. Thus, it only requires adjustments to the site multiplicities or convolution implementation (simplified through, e.g., a Markov chain). In the particular dilute-BCC example at hand, depicted in Figure~\ref{fig:KS2022dilute}, there are 4 such \emph{representation-unique} crystallographic orbits, i.e., 1 with the dilute atom, 2 neighboring the dilute atom sharing either large hexagonal (1st nearest neighbor shell) or small square face (2nd nearest neighbor shell), and 1 non affected by the dilute atom which is equivalent to the remaining 4 orbits; thus reducing number of sites that need to be considered by a factor of 2.

The \texttt{KS2022\_dilute} featurization routine, schematically depicted in Figure~\ref{fig:KS2022dilute}, conveniently automates the above process for both simple cases like aforementioned substitution in pure element and complex cases like introducing a dilute atom at the 2a/2b orbit in $\sigma$-phase (green cell in Fig. \ref{fig:ks2022}), by performing independent identification of crystallographic orbits in the dilute structure and base structure, followed by identification of the dilute site and its configuration to establish orbit equivalency under \texttt{pySIPFENN}'s \texttt{KS2022} representation, to finally reconstruct complete site ensemble of the dilute structure.

In the case of \texttt{KS2022\_dilute} implementation run on the dilute BCC supercell shown in Figure~\ref{fig:KS2022dilute}, the efficiency is improved nearly proportionally to the reduction in the number of considered sites, averaging $51$ms vs $98$ms \texttt{KS2022}, signifying 1.9 computational cost reduction relative to calculating all crystallographically unique sites. Or around \emph{10-fold computational cost reduction} relative to the standard \cite{Ward2017, Chen2019GraphCrystals, Jha2019IRNet, Krajewski2022ExtensibleNetworks, Choudhary2021AtomisticPredictions, Deng2023CHGNetModelling, Davariashtiyani2023FormationRepresentation, Schmidt2023Machine-Learning-AssistedMaterials} approach of processing all sites ($494$ms), while producing precisely the same results (within the numerical precision).

It is also worth mentioning the active interest of the community in this class of atomic structures, as evidenced by the dedicated ADAQ Database (\href{https://defects.anyterial.se}{defects.anyterial.se}) \cite{Davidsson2021ADAQ:Semiconductors} and dedicated ML modeling efforts \cite{Rahman2024AcceleratingNetworks} being published at the time of writing of this work.

\section{Optimizations for Random Solid Solutions} \label{sec:randomsolutions}

Sections \ref{sec:ordered} and \ref{sec:dilute} have demonstrated how recognition of symmetry in ordered structures can guarantee equivalency of sites and how understanding the character of featurization can further extend that notion of equivalency so that the ML representations of all sites can be obtained efficiently up to an order of magnitude faster. Random solid solutions are the conceptually opposite class of atomic structures, where the \emph{lack of} symmetry or site equivalency is \emph{guaranteed}, yet featurizing them requires one to solve the same problem of efficiently obtaining the ML representations of all sites present, which also happen to be infinite.

Typically, in the ab initio community, random solid solutions are represented using Special Quasirandom Structures (SQS) introduced in landmark 1990 work by \citet{Zunger1990SpecialStructures}, which are \emph{the} best structures to match neighborhood correlations in a purely random state given component ratios and number of atoms to use, hence the name \emph{special}. For many years, finding SQS structures required exponentially complex enumeration of all choices and was limited to simple cases until another critical work by \citet{VanDeWalle2013EfficientStructures}, which used simulated annealing Monte Carlo implemented through \texttt{ATAT} software to find these special cases much faster, exemplified through the relatively complex $\sigma$-phase and enabling the creation of SQS libraries used in thermodynamic modeling \cite{vandeWalle2017SoftwareData}.

However, the direct use of an SQS may not be the optimal choice for structure-informed random solid solution featurization due to several reasons. Firstly, as discussed by \citet{VanDeWalle2013EfficientStructures}, SQS can be expected to perform well on purely fundamental grounds for certain properties like total energy calculations, but one has to treat them with caution because different properties will depend differently on the correlation and selecting the SQS may be suboptimal. Building up on that, one could, for instance, imagine a property that depends strongly on the existence of low-frequency, high-correlation regions catalyzing a surface reaction or enabling nucleation of a dislocation. In terms of ML modeling, this notion is taken to the extreme, with calculated features being both very diverse and numerous while being expected to be universal surrogates for such mechanistically complex properties.

Secondly, SQSs that can be generated in a reasonable time are limited in terms of the number of atoms considered, causing quantization of the composition. This is not an issue if a common grid of results is needed, e.g., to fit CALPHAD model parameters \cite{vandeWalle2017SoftwareData} or to train a single-purpose ML model \cite{Tandoc2023MiningAlloys}, but it becomes a critical issue if one needs to accept an arbitrary composition as the ML model and SQS would have to be obtained every time. This issue is further amplified by the rapidly growing field of compositionally complex materials (CCMs), which exist in vast many-component compositional spaces prohibiting SQS reuse even at coarse quantizations \cite{Krajewski2024Nimplex} while being a popular deployment target for both forward and inverse artificial intelligence methods \cite{Catal2023MachineProperties, Rao2022MachineDiscovery, Debnath2023ComparingAlloys} due to their inherent complexity.

Based on the above, it becomes clear that costly computing of an SQS structure would have to be done for every ML model, and it would not be consistent between chemistries and complexities. At the same time, the primary motivation for limiting the number of sites for ab initio calculations is gone since \texttt{KS2022} can featurize over 1,000 sites per second on a laptop (Apple M2 Max run in parallel). 

\begin{figure}[h]
    \centering
    \includegraphics[width=0.95\textwidth]{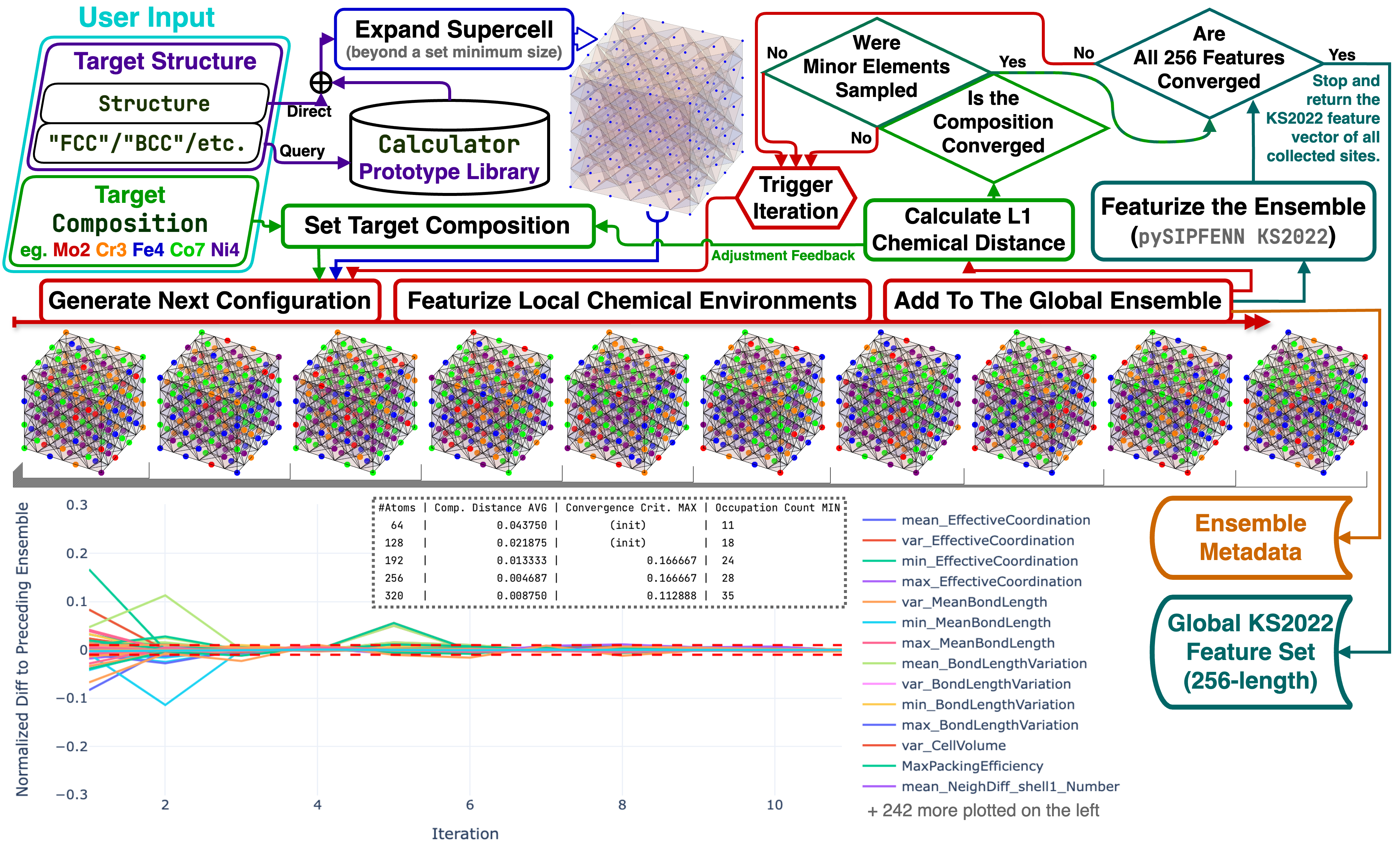}
    \caption{
    Core schematic of the \texttt{KS2022\_randomSolutions} featurizer. The target structure given explicitly or implicitly is expanded to form a (lattice) (i.e. template) supercell. It is then iteratively populated with target composition (slightly adjusted each time) and divided into individual sites, which are featurized (like in \texttt{KS2022}) and added to the global ensemble. The process repeats until the composition is converged, all species have had a chance to occur, and \emph{every individual feature} has converged. Lastly, the global \texttt{KS2022} feature vector and metadata are returned. FCC supercell is used as an example.
    }
    \label{fig:KS2022randomSolution}
\end{figure}

Thus, the objective of optimization is shifted towards consistency in convergence to feature vector values at infinity. To accomplish that, \texttt{pySIPFENN} goes back to random sampling but at a large scale and \emph{individually monitoring the convergence of every feature} during the expansion procedure, implemented through \texttt{KS2022\_randomSolutions} and depicted in Figure~\ref{fig:KS2022randomSolution}, to ensure individual convergence.

Such a representation-centered approach can also efficiently account for (1) the dissimilarity of any set of chemical elements and (2) the neighbor weight during featurization, where some may be much more important than others (see highly-anisotropic $\sigma$-phase Wigner-Seitz cells in Figure~\ref{fig:ks2022}). It is also flexible in accepting any target structure, even a distorted one since no assumptions are made about the neighborhood geometry.

At the same time, it is important to note that such an approach is not a replacement for SQS in a general sense. It is, instead, a complementary method, as it does not result in a defined approximation of random structure but its representation for machine learning.

\section{Summary and Conclusions} \label{sec:summaryconclusions}

\begin{itemize}
    \item \texttt{pySIPFENN} or \textit{python toolset for Structure-Informed Property and Feature Engineering with Neural Networks} is a free open-source software (FOSS) modular framework extending authors' past work \cite{Krajewski2022ExtensibleNetworks} described in Section \ref{sec:Introduction} by including many key improvements in the structure-informed featurization, machine learning model deployment, different types of transfer learning (connected to OPTIMADE API \cite{Evans2024DevelopmentsExchange}), rewrite of key literature tools (e.g., \texttt{Ward2017} Java-based featurizer \cite{Ward2017}) into Python+\texttt{NumPy} \cite{Harris2020ArrayNumPy}, and optimizations of past feature set as described in Sections \ref{ssec:coreimprovements}, \ref{ssec:Ward2017Translation}, and \ref{ssec:ks2022features}.
    
    \item \texttt{pySIPFENN} framework is uniquely built from tightly integrated yet highly independent modules to allow easy use of essential functions without limiting advanced researchers from taking specific components they need, like a specific featurizer, and simply copying it into their software, reducing dependencies to the minimum (including \texttt{pySIPFENN} itself).
    
    \item Section \ref{sec:ordered} discusses how featurization of atomic structures (or configurations) to construct vector, voxel, graph, graphlet, and other representations is typically performed inefficiently because of redundant calculations and how their efficiency could be improved by considering fundamentals of crystallographic (orbits) equivalency to increase throughout of literature machine learning model, typically between 2 to 10 times. Critically, this optimization applies to $98.75\%$ of 4.4 million stored in \texttt{MPDD} \cite{Krajewski2021MPDD:Database}, which includes both DFT-based \cite{Saal2013MaterialsOQMD, Kirklin2015TheEnergies, Shen2022ReflectionsOQMD, Curtarolo2013AFLOW:Discovery, Toher2018TheDiscovery, Jain2013Commentary:Innovation, Choudhary2020TheDesign, Merchant2023ScalingDiscovery} and experimental \cite{Grazulis2009CrystallographyStructures, Grazulis2012CrystallographyCollaboration, Grazulis2019CrystallographyPerspectives} data, showing massive impact if deployed. \texttt{KS2022} featurizer implements these advances in \texttt{pySIPFENN} using \texttt{spglib} \cite{Togo2018Spglib:Search} and \texttt{Voro++} \cite{Rycroft2007MultiscaleFlow, Rycroft2009Voro++:C++, Lu2023AnCells}, while retaining ability to process arbitrary structures.

    \item Section \ref{sec:dilute} explores how symmetry is broken in dilute, doped, and defect structures, to then discuss site equivalency under different representations and how this notion can be used to improve efficiency by skipping redundant calculations of sites which are not guaranteed to be equivalent based on crystallographic symmetry alone but need to be contrasted with defect-free representation. \texttt{KS2022\_dilute} featurizer implements these advances in \texttt{pySIPFENN}.

    \item Section \ref{sec:randomsolutions} discusses featurization of perfectly random configuration of atoms occupying an arbitrary atomic structure and, for the first time, considers fundamental challenges with using SQS approach in the context of forward and inverse machine learning model deployment by extending past discussion on SQS limitations given by \citet{VanDeWalle2013EfficientStructures}, which do not typically appear in ab initio and thermodynamic studies. \texttt{KS2022\_randomSolutions} featurizer 
    has been developed to efficiently featurize solid solutions of any compositional complexity by expanding the local chemical environments (LCEs) ensemble until standardized convergence criteria are met.

    \item As described in Section \ref{sec:softwareavaialbility}, software introduced in this work is continuously tested, well documented, regularly maintained, and 

    \item Throughout this work, the authors explicitly discuss how advances in featurization efficiency described in this work can be applied to different kinds of similar tools in the community, including those using voxel, graph, or graphlet representations.
    
\end{itemize}

\section{Software Availability and Accessibility} \label{sec:softwareavaialbility}

\texttt{pySIPFENN} or \textit{python toolset for Structure-Informed Property and Feature Engineering with Neural Networks} is an easily extensible free, open-source software (FOSS) under \href{https://opensource.org/license/lgpl-3-0}{OSI-approved LGPL-3.0 license}, available as (1) source code hosted in a \texttt{GitHub} repository (\href{https://git.pysipfenn.org/}{git.pysipfenn.org}), (2) a python package through \href{https://pypi.org/project/pysipfenn/}{\texttt{PyPI} index}, and (3) a conda package hosted through \href{https://anaconda.org/conda-forge/pysipfenn}{\texttt{conda-forge} channel}.

It is very well-documented through (1) API reference, (2) detailed changelog, (3) install instructions, (4) tutorials and task-specific notes, and (5) FAQ, compiled for development (\href{https://pysipfenn.org/en/latest/}{pysipfenn.org/en/latest}), stable (\href{https://pysipfenn.org/en/stable/}{pysipfenn.org/en/stable}), and past (e.g., \href{https://pysipfenn.org/en/v0.12.0/}{pysipfenn.org/en/v0.12.0}) versions.

\texttt{pySIPFENN} has been built from the ground up to be a reliable user tool. It is automatically tested across a range of platforms (Linux / Windows / Mac (Intel) / Mac (M1)) and Python versions on every change, as well as on a weekly schedule.

It has been actively disseminated to its target audience through two large workshops organized with support from the Materials Genome Foundation (MGF / \href{https://materialsgenomefoundation.org}{materialsgenomefoundation.org}). The first one, covering \texttt{v0.10.3} and held online on March 2nd 2023, had over 300 users registered and over 100 following all exercises. It has been recorded and published on MGF's YouTube channel \cite{Krajewski20232023YouTube}. The second one, using \texttt{v0.12.1}, was held in-person on June 25th 2023 at the \href{https://calphad.org/calphad-2023}{CALPHAD 2023 conference} in Boston, as a part of Materials Genome Toolkit Workshops, covering its integration with \texttt{ESPEI} \cite{Bocklund2019ESPEICuMg} and \texttt{pycalphad} \cite{Otis2017Pycalphad:Python}. In November 2023, it was also employed in a pair of workshop-style graduate-level guest lectures introducing materials informatics (\href{https://amkrajewski.github.io/MatSE580GuestLectures/}{amkrajewski.github.io/MatSE580GuestLectures}), which can be used as an advanced tutorial.

\section*{Contributions}
\textbf{Adam M. Krajewski:} Conceptualization, Methodology, Software, Investigation, Writing - Original Draft, Validation, Visualization;
\textbf{Jonathan W. Siegel:} Software, Supervision, Writing - Review \& Editing;
\textbf{Zi-Kui Liu:} Funding acquisition, Supervision, Writing - Review \& Editing, Resources

\section*{Acknowledgments}

This work has been funded through grants: U.S. National Science Foundation - Pathways to Enable Open-Source Ecosystems (POSE) \textbf{FAIN-2229690}, DOE Advanced Research Projects Agency-Energy (ARPA-E) \textbf{DE-AR0001435}, and DOE BES (Theoretical Condensed Matter Physics) \textbf{DE-SC0023185}.

We would like to thank (1) \textbf{Jinchao Xu} from PSU/KAUST for his contribution to the development of SIPFENN, published in \cite{Krajewski2022ExtensibleNetworks}; (2) \textbf{Richard Otis} and \textbf{Brandon Bocklund} from \textbf{Materials Genome Foundation} for supporting this work since 2019 in a variety of ways, including invaluable guidance in organizing community workshops; (3) \textbf{Ricardo Amaral} for testing the \texttt{ModelAdjusters} submodule and reviewing the manuscript; (4) \textbf{Rushi Gong, Shuang Lin, ShunLi Shang, Hui Sun, Alexander Richter, and Luke Myers} from Phases Research Lab at PSU, \textbf{Kate Elder} from Lawrence Livermore National Lab, and others for providing feedback when testing the \texttt{pySIPFENN} software; and (5) \textbf{Jan Janssen} from Max-Planck Institut for assistance in deployment and maintenance of \texttt{pySIPFENN}'s \texttt{conda-forge} feedstock.

\printbibliography

\end{document}